\documentclass[a4paper,11pt]{article}
\pdfoutput=1 

\usepackage{jheppub} 

\usepackage[T1]{fontenc} 

\newcommand\blfootnote[1]{%
  \begingroup
  \renewcommand\thefootnote{}\footnote{#1}%
  \addtocounter{footnote}{-1}%
  \endgroup
}

\title{Scanning Strategies at the Top Threshold at ILC \blfootnote{Talk presented at the International Workshop on Future Linear Colliders (LCWS2018), Arlington, Texas, 22-26 October 2018. C18-10-22.}}

\author{Frank Simon}
\affiliation{Max-Planck-Institut f\"ur Physik, Munich, Germany}

\emailAdd{fsimon@mpp.mpg.de}

\abstract{A scan of the top quark pair production threshold at a future electron-positron collider provides the possibility for high-precision measurements of the top quark mass, and, when using two dimensional fits of the measured cross sections, also of other properties such as the width and the Yukawa coupling. The energy range of the scan and the distribution of the integrated luminosity can be optimized depending on the main goals of the threshold program. This contribution examines the possibility to determine the top quark mass in fast exploratory measurements with an adequate precision to enable such an optimization, and studies a scanning program with a reduced energy range of 6 GeV for the measurement of the mass, width and the Yukawa coupling, taking theoretical uncertainties from QCD scale variations and parametric uncertainties from the strong coupling constant into account.}

\notoc

\begin{document} 
\maketitle

\flushbottom

\newpage

\section{Introduction}
\label{sec:intro}

The precise study of top quarks is a central aspects of the physics program of future energy-frontier electron positron colliders. One component the top physics program at such a collider is a scan of the pair production threshold in $e^+e^-$ annihilation, which provides the possibility for a precise measurement of the mass and width, as already recognized well before the discovery of the top quark itself  \cite{Fadin:1987wz, Fadin:1988fn, Strassler:1990nw}.

Previous studies \cite{Martinez:2002st, Seidel:2013sqa, Horiguchi:2013wra} have demonstrated that statistical uncertainties on the level of 20 MeV or better are achievable for the top quark mass with a theoretically well-defined mass definition in a threshold scan with an integrated luminosity of 100 fb$^{-1}$. The main experimental systematics are expected to be on the level of 30 to 50 MeV \cite{Seidel:2013sqa, Simon:2014hna}, with uncertainties from non-resonant top production \cite{Fuster:2015jva} and parametric uncertainties from the strong coupling constant \cite{Simon:2016htt} taking the present uncertainty of the $\alpha_s$ world average contributing at the few 10 MeV level. Theoretical uncertainties, evaluated from scale variation of fixed-order NNNLO QCD calculations \cite{Beneke:2015kwa} in the potential-subtracted (PS) mass scheme \cite{Beneke:1998rk}, have been found to be on the order of 40 MeV \cite{Simon:2016htt, Simon:2016pwp}, putting total uncertainties of 75 MeV or below into reach.

In this contribution, the analysis described in \cite{Seidel:2013sqa, Simon:2016htt, Simon:2016pwp} is extended to two-dimensional extractions of the top quark mass and width and of the mass and the Yukawa coupling. In this framework, the energy range and the scanning strategy of the threshold scan is investigated. Here, the 500 GeV (TDR) version of the International Linear Collider ILC \cite{Behnke:2013xla} is used as the basis for the study, assuming a total integrated luminosity of 200 fb$^{-1}$ used for the threshold scan. Beam polarisation, which would result in more favourable signal to background ratios for certain configurations, is ignored in the present study. The conclusions are in general also valid for CLIC \cite{Lebrun:2012hj, Charles:2018vfv} and FCCee \cite{Benedikt:2651299}. A corresponding study for CLIC assuming a ten-point threshold scan with an integrated luminosity of \mbox{100 fb$^{-1}$}, using the same framework as the study presented here, is included in \cite{Abramowicz:2018rjq}.

\section{The simulation study}

The present study builds on the technique developed in \cite{Seidel:2013sqa}, obtaining simulated cross section measurements from a combination of the NNNLO top pair production cross section with the signal reconstruction efficiencies and expected background levels from full detector simulations performed slightly above the top pair production threshold.  

Following \cite{Seidel:2013sqa}, a signal efficiency of  70.2\% is assumed for the cross section measurement, with an effective background level after all selection cuts of 73 fb. Both numbers are assumed to be energy independent over the range of the threshold scan. The top pair production cross section in the threshold region is generated using version 1.0 of the code QQbar\_threshold \cite{Beneke:2016kkb}, incorporating  fixed-order NNNLO QCD calculations \cite{Beneke:2015kwa} with additional corrections \cite{Beneke:2015lwa, Beneke:2013kia, Beneke:2010mp}.  Throughout the study, a top quark mass of $m_{\mathrm{t}}^{\mathrm{PS}}$ = 171.5 GeV, a width of \mbox{$\Gamma_t$ = 1.37 GeV} and a strong coupling of $\alpha_s(m_{\mathrm{Z}})$ = 0.1185 is assumed. The default scale for the calculation is taken to be $\mu$ = 80 GeV. Within the chosen parameter set, the top quark mass corresponds to a pole mass of $m_{\mathrm{t}}^{\mathrm{pole}}$ = 173.3 GeV, compatible with current LHC measurements. The theory cross section is corrected for initial state radiation and for the luminosity spectrum of the collider as described in \cite{Seidel:2013sqa}.

\begin{figure}
\centering
\includegraphics[width=0.495\textwidth]{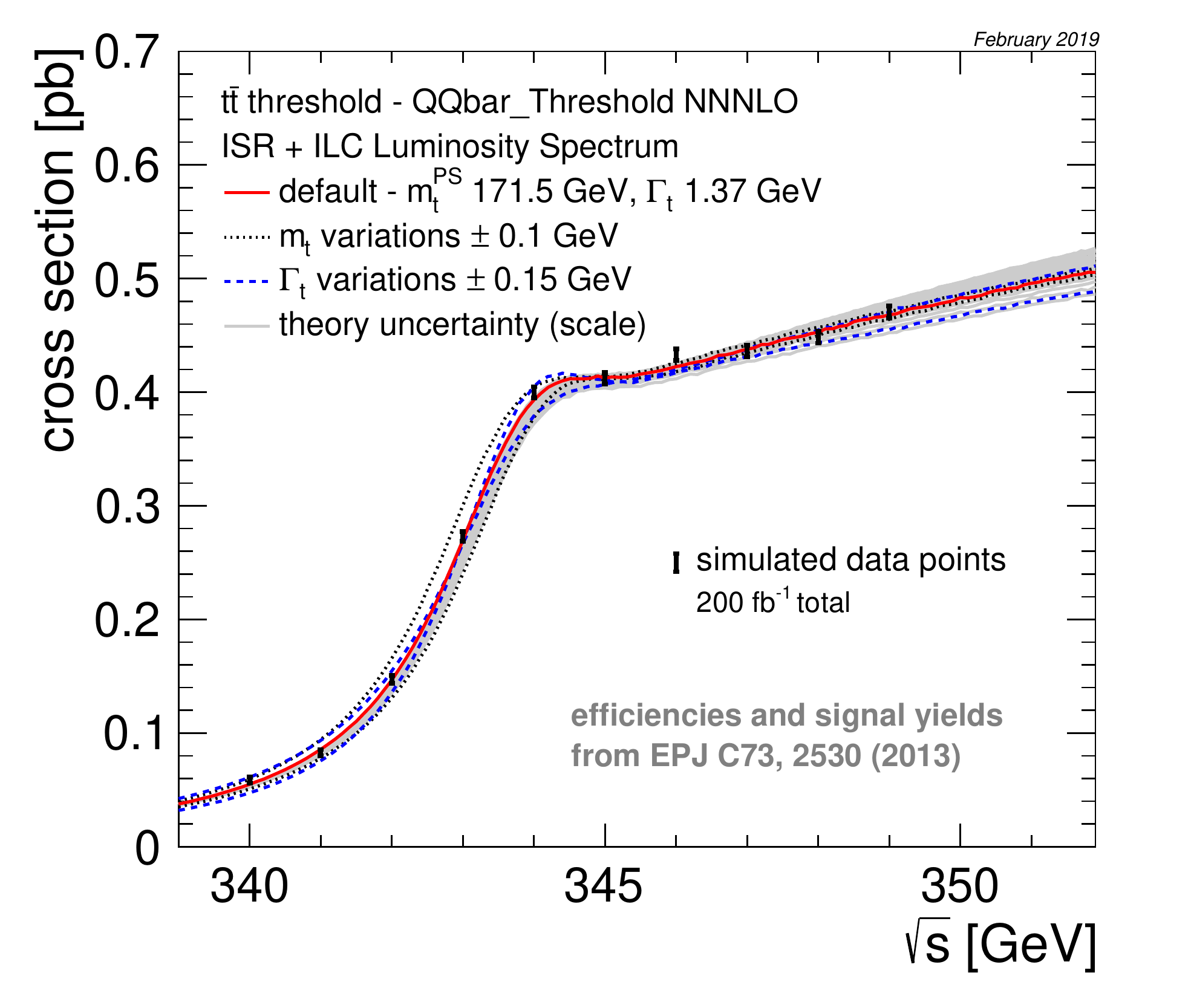}
\hfill
\includegraphics[width=0.495\textwidth]{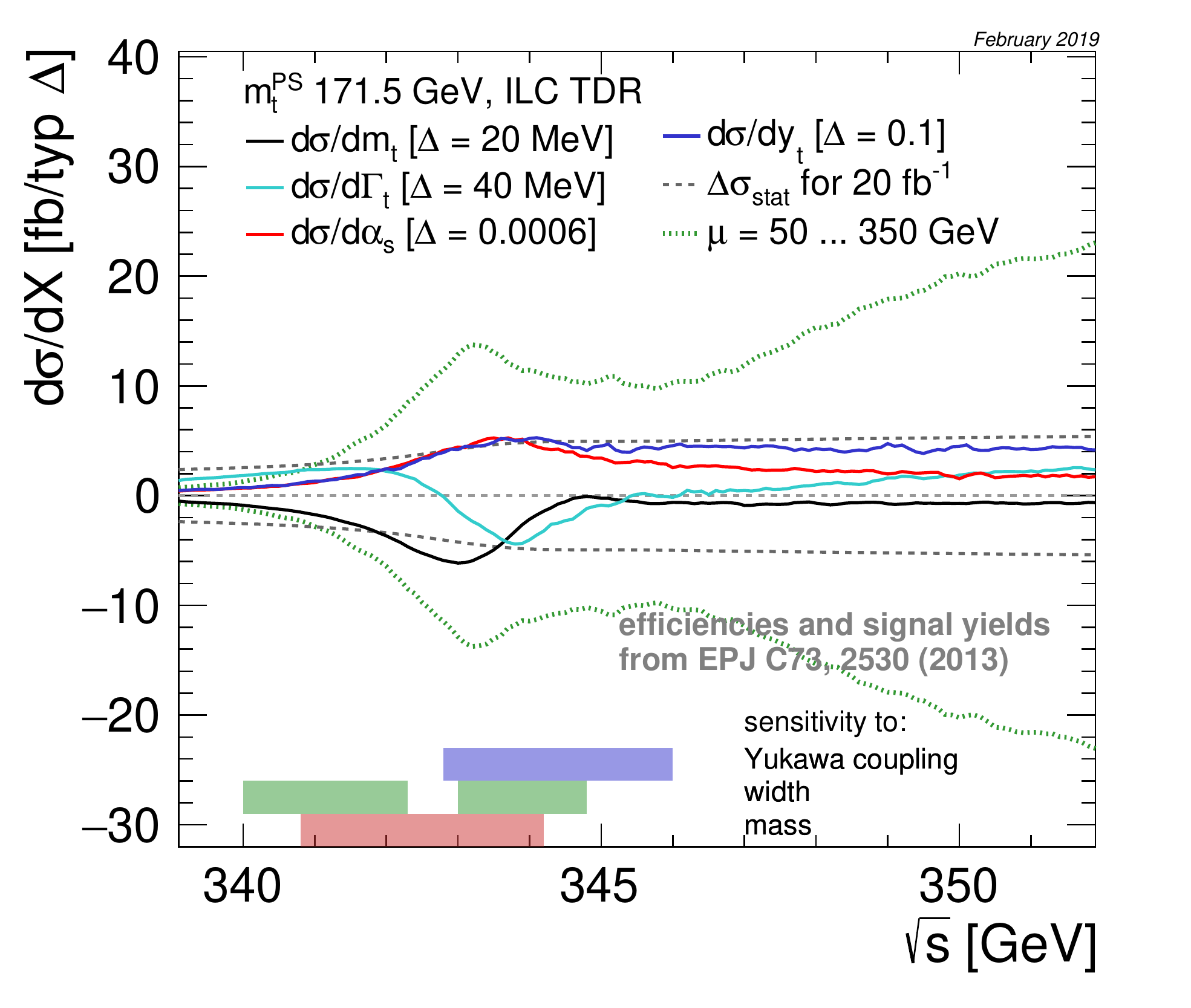}
\caption{{\it Left:} The top pair production cross section in the threshold region, corrected for ISR and the ILC luminosity spectrum. The effect of  mass and width variations as well as the theory uncertainties given by scale variations, and the statistical uncertainty of typical data points with an integrated luminosity of 20 fb$^{-1}$ each, are shown. {\it Right:} The size of the cross section uncertainty due to scale variations compared to variations of the cross section for typical expected statistical uncertainties of measurements at a future collider or external uncertainties for a selection of parameters. Also shown are the statistical uncertainties of a 20 fb$^{-1}$ data point, and the regions most sensitive to particular parameters. \label{fig:CrossSection}}
\end{figure}

Figure \ref{fig:CrossSection} shows the top quark pair production cross section in the threshold region, taking into account initial state radiation and the influence of the ILC luminosity spectrum. The size of the theory uncertainties originating from QCD scale variations relative to typical expected statistical uncertainties of measurements at future colliders or external uncertainties for several parameters influencing the threshold behaviour are also shown. 

The expected statistical uncertainty for single parameter measurements are obtained from the variance of the results of template fits to 10\,000 toy MC experiments using the efficiencies and background levels given above, assuming integrated luminosities as described below. For two dimensional extractions, 1\,000\,000 toy MC experiments are performed, with uncertainties extracted from the 1 $\sigma$ contours of the probability distribution of the fit results. The systematic uncertainties due to scale variations are evaluated by varying the scale used to calculate the input cross section.

\section{Optimising the scan range}
\label{sec:Optimisation}

Figure \ref{fig:CrossSection} {\it right} clearly shows that the sensitivity to the mass $m_t$, width $\Gamma_t$ and Yukawa coupling $y_t$ varies across the threshold region. Depending on the primary goals of the threshold program, the extension in energy and the exact placement of the data points can thus be optimised to minimise the required running time or to maximise the statistical precision. While the highest sensitivity to the mass is given in region of the steepest rise of the cross section, here around 343 GeV, the regions below and slightly above this value are most sensitive to $\Gamma_t$, while sensitivity to $y_t$ is largest at higher energies. 

For an optimal use of the changing sensitivities, the target position of the scan points should be known with an accuracy on the order of 200 - 300 MeV, as also apparent from Figure \ref{fig:CrossSection} {\it right}. This in turn requires a prior knowledge of the top quark mass with a precision of 100 - 150 MeV, since an uncertainty of the mass directly translates into an uncertainty of the exact location of the threshold in energy by a simple multiplication of two. This precision is required in a theoretically well-defined mass scheme, such as the PS mass used here, or the pole mass. Given the uncertainties associated with the interpretation of the mass parameter in event generators used in most top quark mass measurements at the LHC, LHC measurements alone will most likely not reach a sufficient precision. More precise measurements could be obtained from measurements using radiative $t\bar{t}$ events at $e^+e^-$ collision energies in the continuum above the threshold, which may be performed before a threshold scan is performed, as discussed in \cite{Abramowicz:2018rjq}. 

The threshold region itself also offers the possibility to determine the mass in a first initial measurement with sufficient precision with a small fraction of the total integrated luminosity foreseen for a full threshold scan. Assuming an initial uncertainty of the top quark mass in the PS mass scheme of 1 GeV, a single measurement is however not sufficient, since this would require a 4 GeV wide region that is guaranteed to provide unambiguous sensitivity to the mass also when considering theoretical uncertainties. As shown in Figure \ref{fig:CrossSection}, the region with mass sensitivity is indeed approximately 4 GeV wide, but the theoretical uncertainties lead to ambiguities in the region of the would-be 1S peak. This issue can be resolved by performing a first exploratory measurement with two data points of 5 fb$^{-1}$ each, where the data points are placed at $2 \times m_{t, \mathrm{initial}}^{PS} - 1.5\ \mathrm{GeV}$ and at  $2 \times m_{t, \mathrm{initial}}^{PS} + 0.5 \ \mathrm{GeV}$, with $m_{t, \mathrm{initial}}^{PS}$ representing the input top quark mass in the PS mass scheme prior to the threshold scan. With an input uncertainty of 1 GeV, the top quark mass can be determined with a precision of approximately 100 MeV taking the combined statistical and theoretical uncertainties into account. For the purpose of determining the location of the scan points, experimental systematics as well as parametric uncertainties due to the strong coupling constant are not relevant since they are present in the same way in the full scan and thus cancel in this case. 

For the measurement of the top quark mass alone, assuming all other parameters follow the Standard Model, the optimal point for the measurement is the region of the largest derivative of the cross section as a function of the mass, at 343 GeV in the chosen parameter set. Here an integrated luminosity of 200 fb$^{-1}$ concentrated in this one point would result in a statistical uncertainty of approximately 5 MeV. The expected systematic uncertainties far exceed this value, with theoretical uncertainties given by the scale uncertainties of \mbox{41 MeV}, and the parametric uncertainties from the strong coupling constant of 3 MeV per 10$^{-4}$ uncertainty. Considering that such a single point measurement also excludes the measurement of other parameters, it clearly is not a viable option for a top threshold physics program. 

\begin{figure}
\centering
\includegraphics[width=0.495\textwidth]{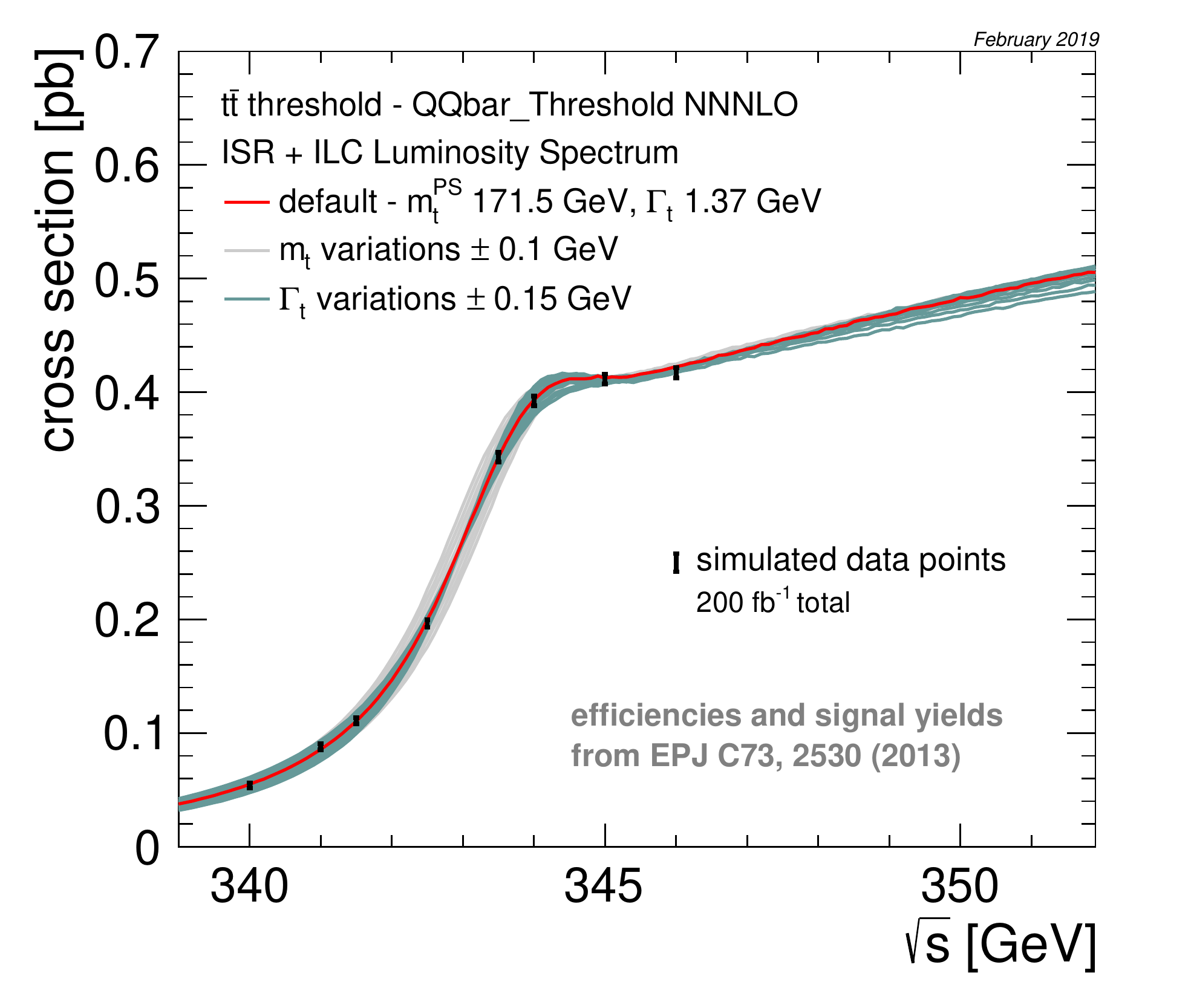}
\caption{Illustration of the 8 point scan covering an energy range of 6 GeV around the threshold, optimised to measure the mass and width of the top quark. The impact of variations of $m_t$ and $\Gamma_t$ are also indicated, showing the placement of the data points in the relevant region. \label{fig:8PointScan}}
\end{figure}

Previous studies have typically assumed a threshold scan with 10 equidistant points spaced by 1 GeV, as illustrated in Figure \ref{fig:CrossSection} {\it left}. In this approach, half of the integrated luminosity is placed in a region that is primarily sensitive to the Yukawa coupling and to the strong coupling constant, and is also characterised by growing theoretical uncertainties. A measurement of $y_t$ at threshold will thus quickly be systematics limited, and not be competitive with other model-dependent indirect measurements and with direct measurements of $t\bar{t}H$ final states. While a certain degree of sensitivity to $y_t$ from the threshold program is interesting, the choice of the energy points should consequently focus more on $m_t$ and $\Gamma_t$. We thus propose a reduced range of 6 GeV, with the measurement points placed preferentially in regions that have sensitivity to both  $m_t$ and $\Gamma_t$. This results in an 8 point scan with 25 fb$^{-1}$ per data point, with the data points placed at 340, 341, 341.5, 342.5, 343.5, 344, 345 and 346 GeV, as illustrated in Figure \ref{fig:8PointScan}, which also shows the variations of the cross section with $m_t$ and $\Gamma_t$. A numerical optimisation of the location of these scan points has not been carried out, the proposal represents an ad-hoc choice based on the arguments outlined above. In general, the performance does not depend strongly on the exact placement of the data points, as long as the regions with sensitivity to the relevant parameters are adequately covered.

\section{Results with an 8 point scan}

The top quark pair production threshold scan introduced in Section \ref{sec:Optimisation} has been studied for the ILC TDR luminosity spectrum, assuming a total integrated luminosity of 200 fb$^{-1}$. The results are compared to the scenario used in previous studies, with the same total integrated luminosity spread over 10 equidistant energy points from 340 GeV to 349 GeV. 

\subsection{Measurement of the top quark mass}

For a one dimensional fit of $m_t$, the statistical uncertainty is 10.3 MeV, compared to \mbox{12.2 MeV} for the 10 point scan. At the same time, the theoretical uncertainty originating from scale uncertainty increases to 43 MeV from 40 MeV with the reduced range of the threshold scan. The parametric uncertainty due to the strong coupling constant remains unchanged at 2.9 MeV per 10$^{-4}$ uncertainty in $\alpha_s$.

\subsection{Two-dimensional fits of cross section measurements in the threshold region}

\begin{figure}
\centering
\includegraphics[width=0.495\textwidth]{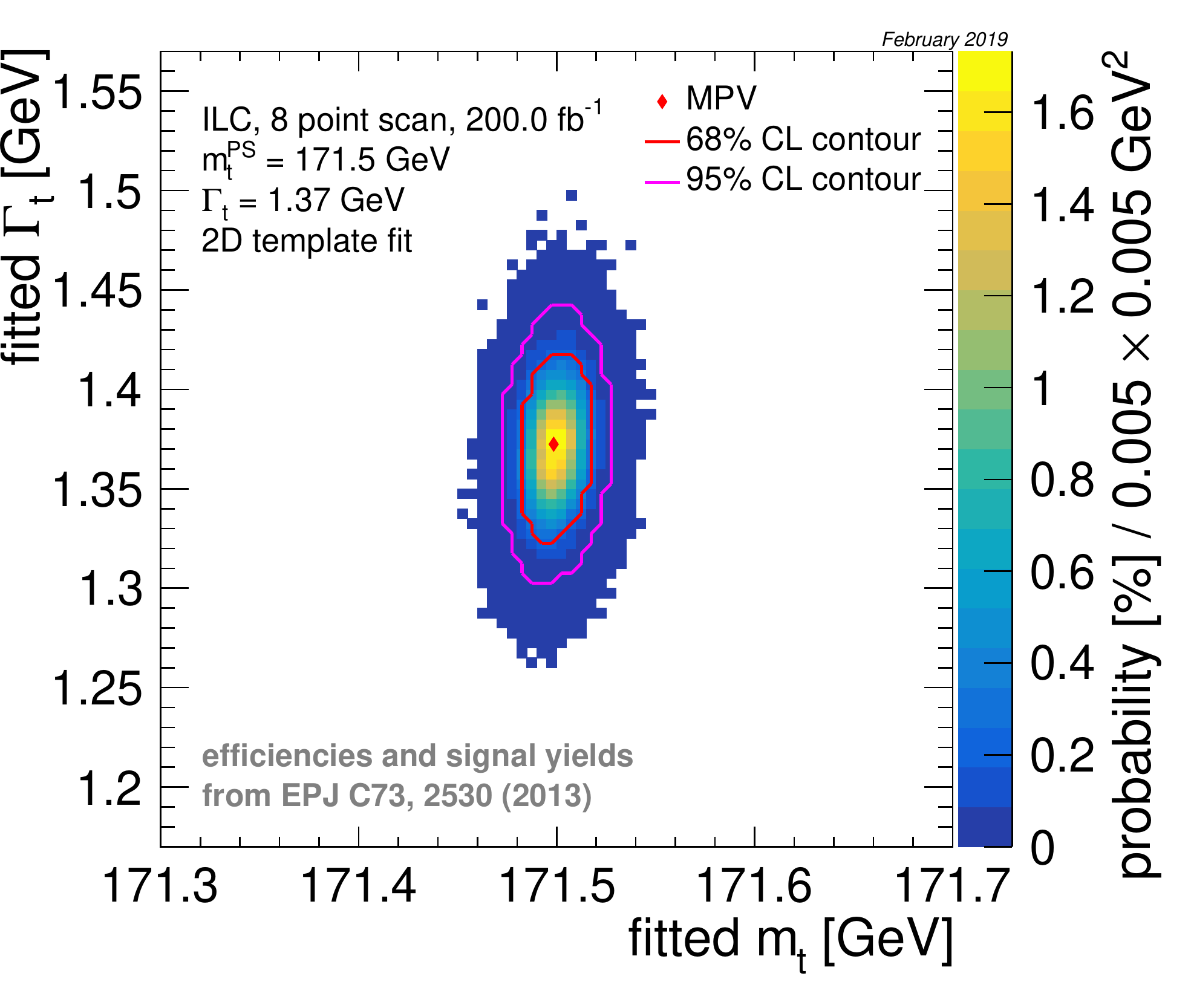}
\hfill
\includegraphics[width=0.495\textwidth]{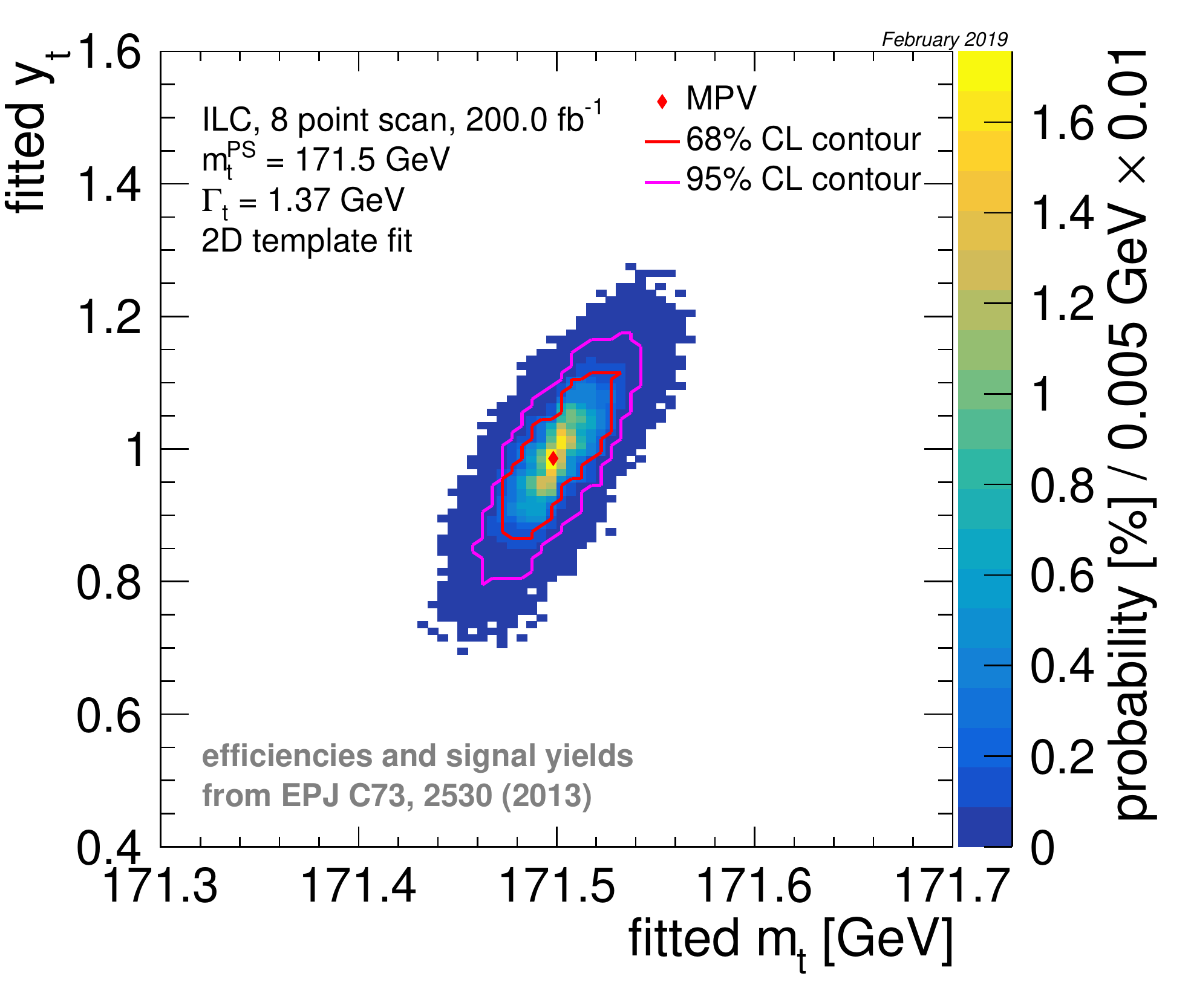}
\caption{The probability distribution of fitted $\Gamma_t$ and $m_t$ ({\it left}) and $y_t$ and $m_t$ ({\it right}) values in two-dimensional fits of the top threshold using the 8 point scan at ILC, obtained from one million toy MC experiments each. The 68\%  and 95\% CL $\sigma$ contours are also shown.\label{fig:2D}}
\end{figure}

The capability for simultaneous measurement of two parameters of the top quark, either $\Gamma_t$ and $m_t$ or $y_t$ and $m_t$, is studied by performing a two-dimensional template fit of simulated cross-section measurements in the threshold region. The fit is performed by calculating the $\chi^2$ of the data points with respect to a two-dimensional grid of cross section templates for different $\Gamma_t$ and $m_t$ or $y_t$ and $m_t$ values, and subsequently fitting a two-dimensional parabola to the $\chi^2$ distribution. The fitted pair of parameters is given by the minimum of this parabola. The statistical uncertainty and the correlation between parameters are studied by performing one million toy experiments. Figure \ref{fig:2D} shows the result of these fits for the 8 point scan. 

\begin{figure}
\centering
\includegraphics[width=0.45\textwidth]{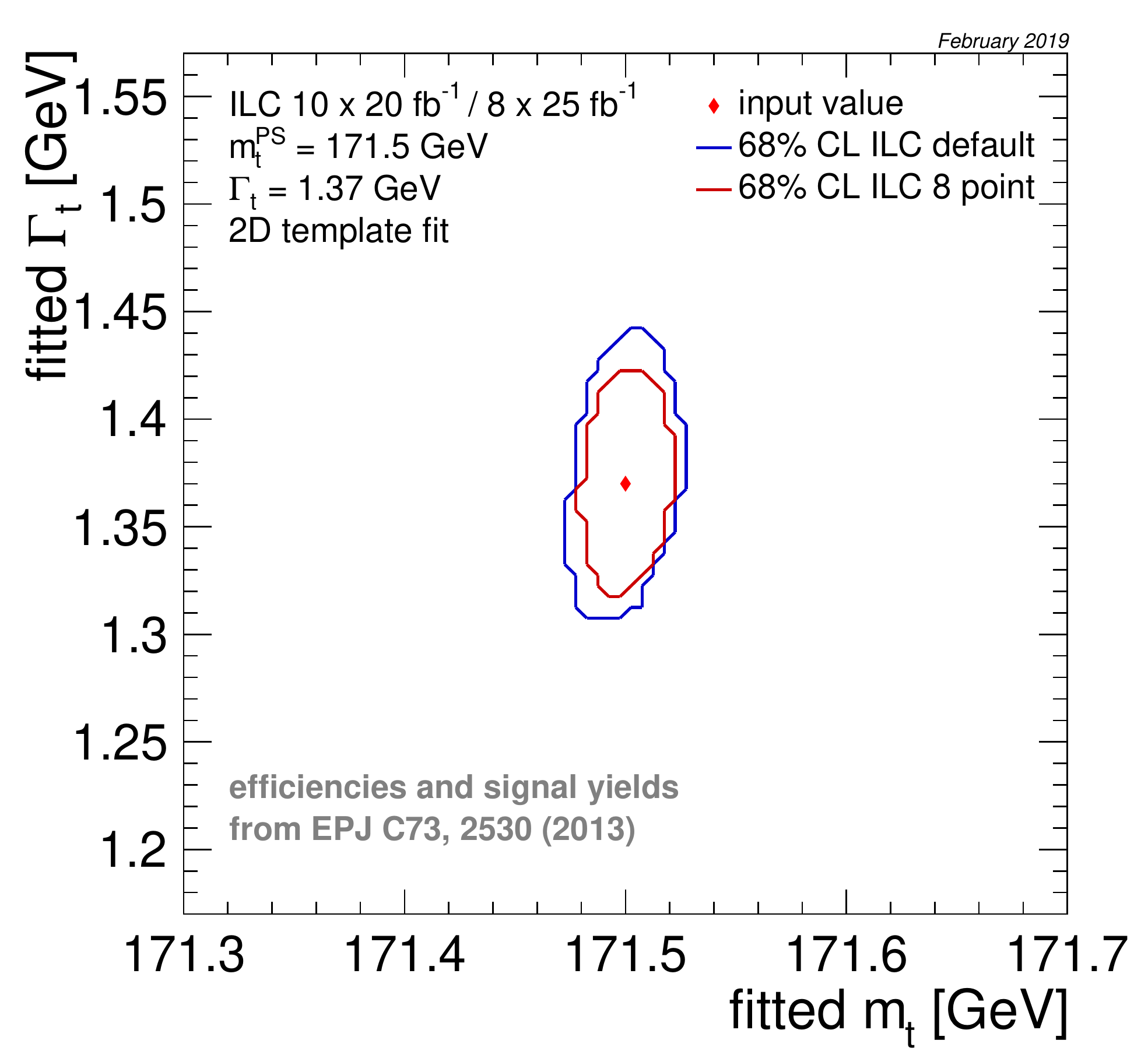}
\hspace{0.01\textwidth}
\includegraphics[width=0.45\textwidth]{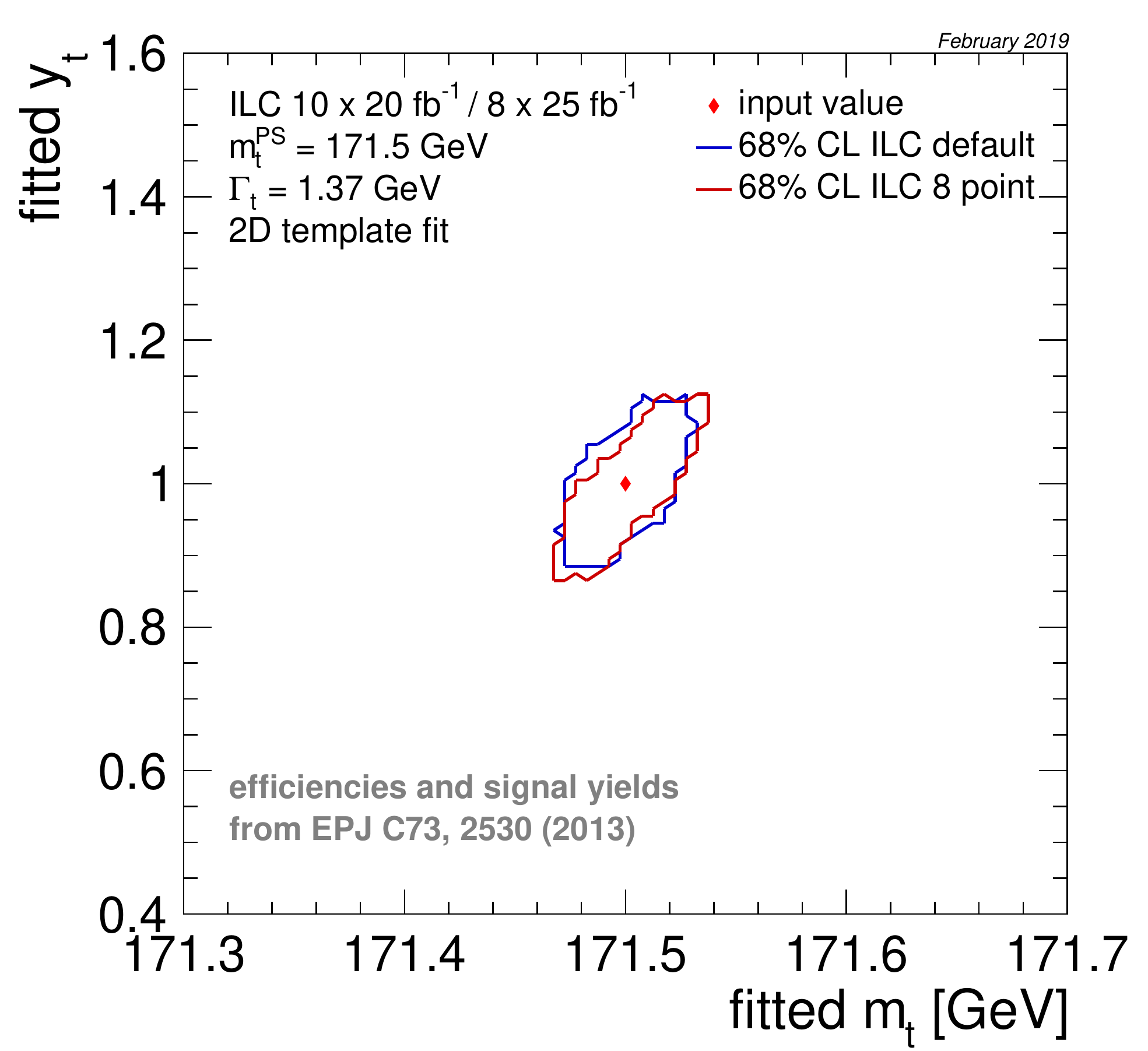}
\caption{Comparison of the 68\% CL contours of two dimensional fits of  $\Gamma_t$ and $m_t$ ({\it left}) and $y_t$ and $m_t$ ({\it right}) for the two different threshold scan scenarios discussed here, the 8 point scan with reduced energy range, and the standard 10 point scan.\label{fig:2DContours}}
\end{figure}

Figure \ref{fig:2DContours} illustrates the impact of changing the threshold scan range on the two-dimensional measurements. While there is a substantial improvement in the fit of $\Gamma_t$ and $m_t$, in particular on $\Gamma_t$, the 8 point fit results in a slight deterioriation of the fit of $y_t$, and an increase in the correlation between $y_t$ and $m_t$. This behavior is expected, since the 8 point scan substantially reduces the amount of integrated luminosity above the threshold, where the sensitivity to $y_t$ is largest and the sensitivity to $m_t$ is low. 

\begin{table}
\centering
\caption{Summary of the results of 1D and 2D fits for the two threshold scan scenarios. For the 2D fits, the statistical uncertainties give the extent of the 68\% CL contour in the respective direction. Also shown is the marginalized 1 $\sigma$ uncertainty, where the same theoretical uncertainties apply. \label{tab:Results}}

\vspace{2ex}
\begin{small}
\begin{tabular}{l|c|c|c|c}
\hline
\hline
parameter & \multicolumn{2}{c|}{8 point scan} & \multicolumn{2}{c}{10 point scan} \\
\hline
\multicolumn{5}{l}{1D fit}\\
\hline
$m_t$ & \multicolumn{2}{c|}{($\pm 10.3${\tiny (stat)} $\pm$ 44{\tiny (theo)}) MeV} & \multicolumn{2}{c}{(12.2{\tiny (stat)} $\pm$ 40{\tiny (theo)}) MeV} \\
\hline
\multicolumn{2}{l|}{2D fit $m_t$ and $\Gamma_t$} & {\small marg.}& & {\small marg.}\\
\hline
$m_t$ & ($^{+20.7}_{-24.3}${\tiny (stat)} $\pm$ 45{\tiny (theo)}) MeV & 10.7 MeV &($^{+29.7}_{-25.3}${\tiny (stat)} $\pm$ 43{\tiny (theo)}) MeV & 13.0 MeV \\
$\Gamma_t$ & ($^{+50}_{-55}${\tiny (stat)} $\pm$ 32{\tiny (theo)}) MeV & 25 MeV &($^{+80}_{-55}${\tiny (stat)} $\pm$ 39{\tiny (theo)}) MeV & 30 MeV\\
\hline
\multicolumn{2}{l|}{2D fit $m_t$ and $y_t$}& {\small marg.} & & {\small marg.}\\
\hline
$m_t$ & ($\pm 35${\tiny (stat)} $\pm$ 45{\tiny (theo)}) MeV & 17.0 MeV &($^{+34}_{-31}${\tiny (stat)} $\pm$ 42{\tiny (theo)}) MeV & 15.2 MeV \\
$y_t$ & $^{+0.120}_{-0.140}${\tiny (stat)} $\pm$ 0.09{\tiny (theo)}  & 0.055 & $^{+0.128}_{-0.112}${\tiny (stat)} $\pm$ 0.132{\tiny (theo)} & 0.047 \\
\hline
\hline
\end{tabular}
\end{small}

\end{table}

Table \ref{tab:Results} summarizes the results for both threshold scan scenarios. For the two dimensional fits the extension of the 68\% CL contour in the corresponding direction, relative to the mean fitted value to capture potential asymmetries of the contours, is given as the uncertainty. Theoretical uncertainties are symmetrized. Also included are the marginalized 1 $\sigma$ uncertainties for each of the parameters in the two dimensional fits, where the same theoretical uncertainties apply. For the mass and the Yukawa coupling, the parametric uncertainty originating from the strong coupling constant has also been evaluated. For $m_t$, this amounts to 2.9 MeV per 10$^{-4}$ uncertainty in $\alpha_s$ for the one dimensional fit, irrespective of the scanning scenario. For the two-dimensional fit it is 1.7 MeV for $m_t$ and 0.0088 for $y_t$ per 10$^{-4}$ uncertainty in $\alpha_s$ for the 8 point scan. For the original 10 point scan, the parametric uncertainty on $y_t$ is slightly reduced to 0.0075 per 10$^{-4}$ uncertainty in $\alpha_s$, while the uncertainty for $m_t$ remains the same. 

\section{Conclusions}

A scan of the top quark pair production threshold at the future ILC or other energy-frontier $e^+e^-$ colliders will provide a highly precise measurement of the top quark mass in theoretically well-defined mass schemes with a precision of 75 MeV or better. Beyond the mass, measurements of the top quark width or of the Yukawa coupling can also be performed by applying two-dimensional fits to the measured cross sections in the threshold region. Here, the optimization of the energy range of the scan, and of the distribution of the integrated luminosity within this range has been studied. Such an optimization implies the need for a prior knowledge of the top quark mass to within 150 MeV in theoretically well-defined mass schemes, which can be achieved with an exploratory measurement using two energy points with 5 fb$^{-1}$ each, spaced by 2 GeV. With an initial precision of the top quark mass of 1 GeV such a measurement will result in an adequate precision to determine the required range and placement of the measurement points of the full scan. Since relatively large theoretical and parametric uncertainties limit the achievable precision on $y_t$, the scanning strategy proposed here concentrates primarily on the region most sensitive to mass and width, resulting in the choice of an 8 point scan spanning 6 GeV from $2 \times m_t^{PS}$ - 3 GeV to \mbox{$2 \times m_t^{PS}$ + 3 GeV}. With such a scan, top quark mass measurements with a precision of better than \mbox{75 MeV}, measurements of the width on the level of better than 100 MeV, and a measurements the Yukawa coupling to 15\% are possible with a total integrated luminosity of 200 fb$^{-1}$, considering both statistical and the leading systematic uncertainties. While the present study has been performed assuming the luminosity spectrum of ILC, the conclusions for other collider options such as CLIC or FCCee are similar.

\bibliographystyle{JHEP}
\bibliography{ttbar}

\end{document}